\documentclass[10pt,journal,compsoc]{IEEEtran}

\usepackage{booktabs} 
\usepackage{xcolor}
\usepackage{bm}
\usepackage{subcaption}
\usepackage{graphicx}
\usepackage{amsmath}
\usepackage{amssymb}

\usepackage{hyperref}
\usepackage{colortbl}
\usepackage[capitalise]{cleveref}
\usepackage[ruled]{algorithm2e} % For algorithms

\SetAlFnt{\small}
\SetAlCapFnt{\small}
\SetAlCapNameFnt{\small}
\SetAlCapHSkip{0pt}

\begin{document}
% Title portion
\definecolor{nicegreen}{rgb}{0.2, 0.8, 0.2}
\newcommand{\Note}[1]{{\color{nicegreen}{\bf NOTE: }#1}}

\title{NAT: Neural Acoustic Transfer for Interactive Scenes in Real Time}

% DO NOT ENTER AUTHOR INFORMATION FOR ANONYMOUS TECHNICAL PAPER SUBMISSIONS TO SIGGRAPH 2019!
%\author{Gang Zhou}
%\orcid{1234-5678-9012-3456}
%\affiliation{%
%  \institution{College of William and Mary}
%  \streetaddress{104 Jamestown Rd}
%  \city{Williamsburg}
%  \state{VA}
%  \postcode{23185}
%  \country{USA}}
%\email{gang_zhou@wm.edu}
%\author{Valerie B\'eranger}
%\affiliation{%
%  \institution{Inria Paris-Rocquencourt}
%  \city{Rocquencourt}
%  \country{France}
%}
%\email{beranger@inria.fr}
%\author{Aparna Patel}
%\affiliation{%
% \institution{Rajiv Gandhi University}
% \streetaddress{Rono-Hills}
% \city{Doimukh}
% \state{Arunachal Pradesh}
% \country{India}}
%\email{aprna_patel@rguhs.ac.in}
%\author{Huifen Chan}
%\affiliation{%
%  \institution{Tsinghua University}
%  \streetaddress{30 Shuangqing Rd}
%  \city{Haidian Qu}
%  \state{Beijing Shi}
%  \country{China}
%}
%\email{chan0345@tsinghua.edu.cn}
%\author{Ting Yan}
%\affiliation{%
%  \institution{Eaton Innovation Center}
%  \city{Prague}
%  \country{Czech Republic}}
%\email{yanting02@gmail.com}
%\author{Tian He}
%\affiliation{%
%  \institution{University of Virginia}
%  \department{School of Engineering}
%  \city{Charlottesville}
%  \state{VA}
%  \postcode{22903}
%  \country{USA}
%}
%\affiliation{%
%  \institution{University of Minnesota}
%  \country{USA}}
%\email{tinghe@uva.edu}
%\author{Chengdu Huang}
%\author{John A. Stankovic}
%\author{Tarek F. Abdelzaher}
%\affiliation{%
%  \institution{University of Virginia}
%  \department{School of Engineering}
%  \city{Charlottesville}
%  \state{VA}
%  \postcode{22903}
%  \country{USA}
%}

%\renewcommand\shortauthors{Zhou, G. et al}

\author{Xutong Jin, Bo Pang, Chenxi Xu, Xinyun Hou, Guoping Wang, Sheng Li*,~\IEEEmembership{Member,~IEEE} 
\IEEEcompsocitemizethanks{
\IEEEcompsocthanksitem  Xutong Jin, Bo Pang, Chenxi Xu, Xinyun Hou, Guoping Wang, and Sheng Li are with the School of Computer Science, Peking University, China.\\ E-mail: \{jinxutong\,$|$pangbo\,$|$2301213239\}@pku.edu.cn, \\
\{houxinyun\}@stu.pku.edu.cn, \\ 
\{wgp\,$|$lisheng\}@pku.edu.cn \\
Guoping Wang and Sheng Li are also with the National Key Laboratory of Intelligent Parallel Technology. \\
\IEEEcompsocthanksitem Sheng Li is the corresponding author.
}
}

\IEEEtitleabstractindextext{
\begin{abstract}

%Prior acoustic transfer methods involve extensive precomputation and storage of acoustic transfer data for objects to facilitate real-time interaction and auditory feedback. However, these methods struggle in more complex scenes, particularly when there are dynamic changes in the position, material, and size of objects, which can significantly alter the sound efffect. The continuous variation in these conditions leads to a fluctuating distribution of acoustic transfer, presenting challenges in representing this variation with basic data structures and efficiently rendering acoustic transfer in real-time.
Previous acoustic transfer methods rely on extensive precomputation and storage of data to enable real-time interaction and auditory feedback. However, these methods struggle with complex scenes, especially when dynamic changes in object position, material, and size significantly alter sound effects. These continuous variations lead to fluctuating acoustic transfer distributions, making it challenging to represent with basic data structures and render efficiently in real time.
To address this challenge, we present Neural Acoustic Transfer, a novel approach that utilizes an implicit neural representation to encode precomputed acoustic transfer and its variations, allowing for real-time prediction of sound fields under varying conditions. 
To efficiently generate the training data required for the neural acoustic field, we developed a fast Monte-Carlo-based boundary element method (BEM) approximation for general scenarios with smooth Neumann conditions. Additionally, we implemented a GPU-accelerated version of standard BEM for scenarios requiring higher precision. These methods provide the necessary training data, enabling our neural network to accurately model the sound radiation space.
%To train our neural network efficiently, we implemented two fast BEM solvers for different use cases, providing the training data essential for our neural network to accurately model the entire sound field.
We demonstrate our method's numerical accuracy and runtime efficiency (within several milliseconds for 30s audio) through comprehensive validation and comparisons in diverse acoustic transfer scenarios. Our approach allows for efficient and accurate modeling of sound behavior in dynamically changing environments, which can benefit a wide range of interactive applications such as virtual reality, augmented reality, and advanced audio production. 

% Acoustic Transfer在虚拟场景中的声音合成问题里是非常重要的一个环节，为了能够在runtime与一个虚拟物体实时交互并听到它发出的声音，需要预计算并存储物体的acoustic transfer。但是对于更复杂的场景，物体的位置，材质，大小等会发生变化而影响声音，为了将各种情况的声场分布预计算并存储下来，之前的方法不再适用。因为在各种条件的连续变化下，声场分布也会随之连续变化。这种由复杂条件张成的高维空间很难用简单的数据结构准确表示。为了解决这个问题，我们提出了 Neural Acoustic Transfer, 一种用隐式神经网络来编码acoustic transfer的预计算方法，可以实时预测不同条件下的声场。为了快速有效地训练我们的Neural Acoustic Transfer，我们首先在整个条件空间随机采样条件参数，在每个条件参数下，对场景表面均匀随机采样，再用基于蒙特卡洛积分的解法对这些表面采样点的声压值进行快速估计，并用这些采样点的估计值来训练神经网络拟合整个声场。We demonstrate the numerical accuracy, the runtime performance of our method on a set of comparisons and examples

\end{abstract}

}

\maketitle
\IEEEpeerreviewmaketitle

\begin{figure*}
  \centering
  \begin{subfigure}[b]{0.49\linewidth}
    \includegraphics[trim={0cm, 0cm, 0cm, 0cm},clip,width=\linewidth]{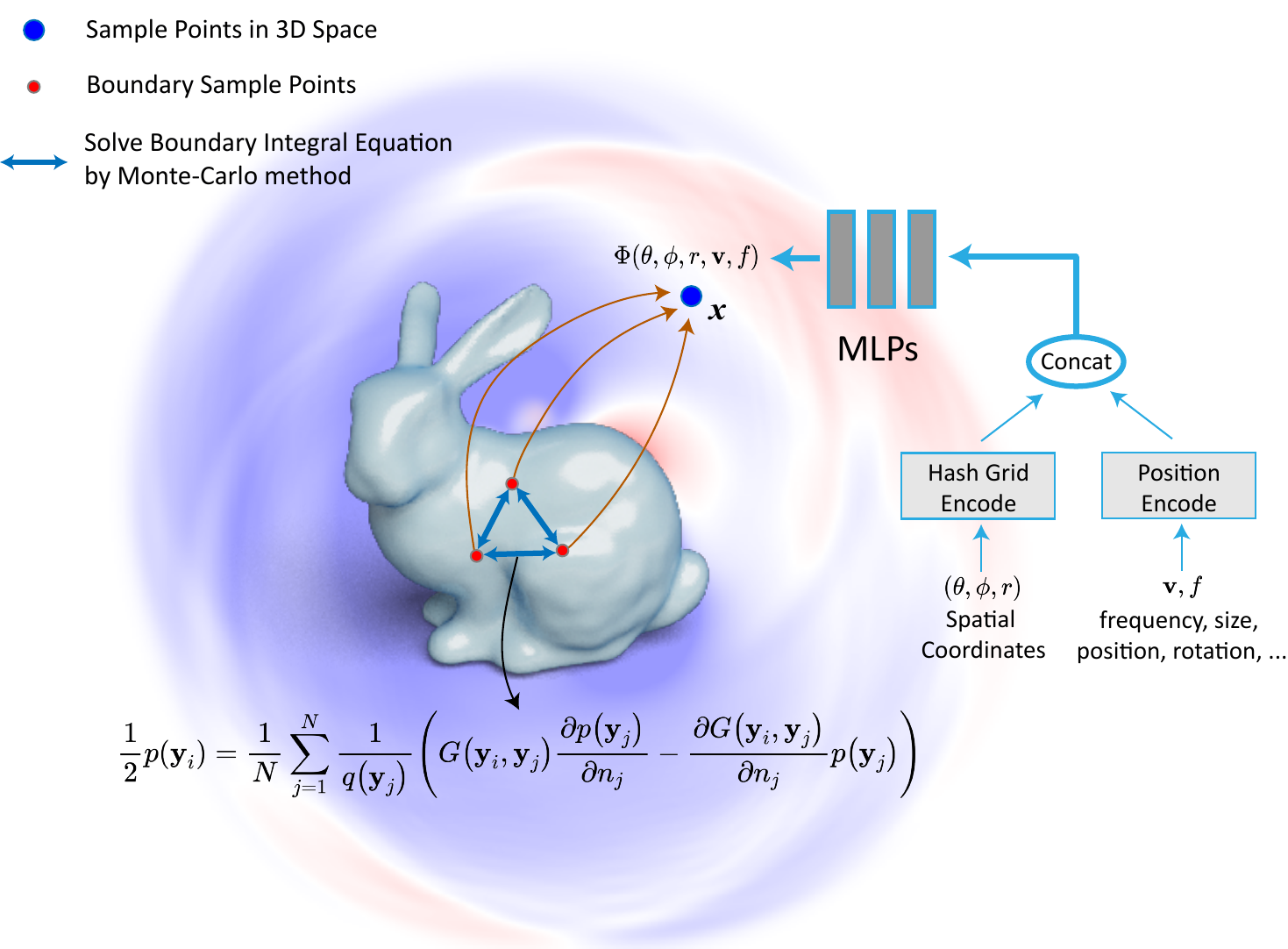}
    \caption{Our neural representation handles the complex acoustic transfer function, using data from a Monte-Carlo-based solution or classical numerical solver.}
    \label{fig:overview}
  \end{subfigure}
  \qquad
%  \hfill % this will add a little space between the two subfigures
  \begin{subfigure}[b]{0.46\linewidth}
    \includegraphics[trim={0cm, 0.5cm, 0cm, 0cm},clip,width=\linewidth]{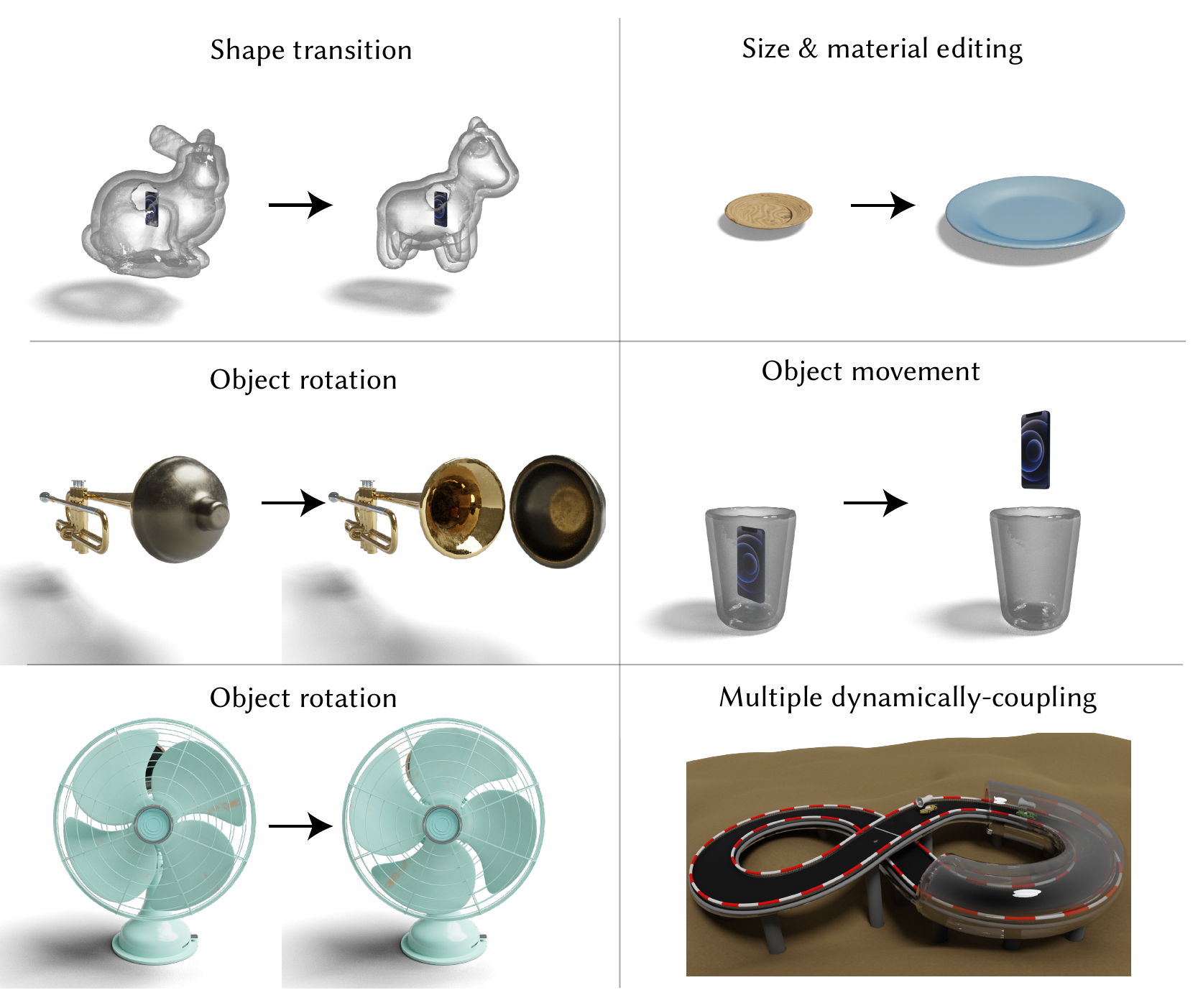}
    \caption{Our approach can handle various dynamic scenarios for real-time acoustic transfer prediction.}
    \label{fig:teaser}
  \end{subfigure}
  \caption{
Our Neural Acoustic Transfer implicitly models the acoustic transfer function under changing conditions, as well as a Monte-Carlo-based approach to synthesize the training data efficiently. Our approach demonstrates real-time performance, achieving sound simulation within 3 milliseconds (for audio lengths of 30 seconds) in various dynamically changing scenarios with exceptional accuracy.}
  \label{fig:teaserfigure}
\end{figure*}

\section{Introduction}

Acoustic transfer plays a crucial role in synthesizing plausible, physically-based sounds synchronized with computer-simulated animations~\cite{pai2001scanning, o2002synthesizing, PAT,shell-ffat}. Typically, physical simulation methods are capable of computing the acceleration of object surfaces, as seen in simulations involving rigid bodies and fluids. These accelerations then serve as inputs to acoustic transfer solvers, which estimate the sound pressure at any point in space. For scenarios involving a single solid object, precomputed methods~\cite{PAT, shell-ffat} rely on offline solvers, often boundary element methods (BEM)~\cite{bem_survey}, to precompute and store the sound field distribution produced by the modal vibrations of an object, enabling frequent runtime evaluation.

However, these methods have difficulties in handling more complex scenes. For instance, Li et al.~\cite{li2015interactive} proposed precomputed methods that allow for interactive and continuous editing and exploration of modal sound parameters. Their methods still struggle to achieve real-time performance. In dynamic scenes, where sound-emitting objects move relative to other scene elements, the sound field continuously changes. Real-time interactivity in such complex scenes requires precomputing and storing acoustic transfer distributions for all possible situations. Representing these distributions accurately in a high-dimensional space poses a challenge, while previous methods are incapable of handling these complex tasks efficiently.

Considering implicit representation by neural networks is known for its robust representational capabilities and rapid inference in various domains~\cite{nerf, neural_radiosity}, we present the Neural Acoustic Transfer (NAT) approach. Our approach utilizes neural networks to encode precomputed acoustic transfers within a high-dimensional space, enabling real-time interaction in complex dynamic scenes. Spatial coordinates of listener positions are encoded with a multi-resolution hash grid~\cite{mueller2022instant, tiny-cuda-nn}, and other condition parameters are encoded using positional encoding~\cite{nerf}. These encoded inputs are then concatenated and processed by the neural network to fit the acoustic transfer in the high-dimensional space using data obtained from a numerical solver.
To facilitate synthesizing the training data, we implemente a CUDA-accelerated Boundary Element Method (BEM) for modal sound acoustic transfer simulation and propose a fast Monte-Carlo-based approximation strategy for BEM in scenes with smoother boundary conditions. This Monte-Carlo-based approximation is less sensitive to mesh quality compared to traditional BEM and can significantly reduce computational costs by controlling the sample count. However, it achieves moderate accuracy, which is sufficient for sound effect purposes.

Through comprehensive experiments, we validate the effectiveness and superiority of our Neural Acoustic Transfer method in dynamic scenes. This includes editing material properties and sizes for modal sound objects, as well as rendering dynamic acoustic effects within motion-coupled environments (see \cref{fig:teaserfigure}). Remarkably, our approach can predict sound variations in dynamic scenes for a single listener position within just \(1 \, \text{ms}\) for audio of length \(10 \, \text{s}\). For material and size editing of a modal sound object, our approach can predict the acoustic transfer map of the modal sound object for new material sizes in \(2 \, \text{ms}\), achieving speedups of several orders of magnitude over the previous neural approach (NeuralSound \cite{NeuralSound}) and with higher precision. This breakthrough makes real-time acoustic interaction in complex dynamic scenes a feasible and practical reality.

Overall, our main contributions include:
\begin{itemize}
    \item We present a neural network to precisely represent the complex acoustic transfer distribution across dynamically changing scene conditions. The dynamic sound effect with acoustic transfer can be simulated in real time.
    
    \item We develop a Monte-Carlo-based approximation for BEM, which can be used to synthesize acoustic transfer datasets efficiently for neural network training.
    
   \item We provide the practical applications of Neural Acoustic Transfer. These include real-time editing of material properties and sizes for modal sound objects, as well as the rendering of dynamic acoustic effects.
   
\end{itemize}

\section{Related work}

\subsection{Acoustic Transfer}

Acoustic transfer modeling is a crucial aspect of simulating how vibrations on an object's surface translate into sound perceivable by the human ear. Traditionally, the Boundary Element Method (BEM)~\cite{bem_survey} has been a go-to technique for its high accuracy, often serving as a de facto standard or `ground truth' in acoustic simulations. However, BEM's major drawback lies in its computational intensity, rendering it less viable for dynamic or real-time scenarios. To circumvent this limitation, Precomputed Acoustic Transfer (PAT)~\cite{PAT} methods emerged, leveraging BEM's precomputed data. These methods approximate the acoustic field around a vibrating object using a set of simpler sound sources, significantly accelerating the inference process. Another approach is the utilization of single or multiple sound sources to represent the acoustic field equivalently~\cite{contact,fracture,eigenmode-compression,rungta2016syncopation}. The Far-Field Acoustic Transfer (FFAT) Maps~\cite{shell-ffat,kleinpat}, offer a different perspective by encoding the sound field on a spherical surface surrounding the object, unfolding it into an image-like representation. Our Monte-Carlo-based approximation for BEM provides an advanced alternative to traditional BEM, maintaining high accuracy without being hindered by mesh quality. Meanwhile, our Neural Acoustic Transfer technique stands as a novel representation in acoustic modeling, outperforming traditional methods like PAT and FFAT maps in adaptability for complex dynamic scenes.

\subsection{Neural field with sound simulation}
The emergence of neural fields has revolutionized signal representation by offering high fidelity reconstruction~\cite{mildenhall2021, sitzmann2020, tancik2020}, enabling sampling at arbitrary locations, and providing fast training and inference~\cite{chan2022, chen2022b, fridovich2022}. Primarily applied in image compression~\cite{martel2021}, view synthesis~\cite{barron2021, mildenhall2021, verbin2022}, 3D reconstruction~\cite{mescheder2019, park2019, peng2020, wang2021}, and generation~\cite{cai2020, chen2020, chen2019learning, mescheder2019, park2019, yang2019}, neural fields have shown potential in various domains. Sitzmann et al.~\cite{sitzmann2020} demonstrated their use with specific architectures to solve PDEs, inspiring applications in geometry processing tasks~\cite{yang2021a}, character animation~\cite{bergman2022, noguchi2021}, level-set methods~\cite{mehta2022}, and PDE solving~\cite{chen2022a, li2023}. In the field of sound simulation, recent deep learning methods have been proposed for sound synthesis~\cite{hawley2020, ji2023, jin2020, NeuralSound}, scattering effect computation, and sound propagation~\cite{fan2020, pulkki2019, ratnarajah2020, tang2021, luo2022learning}, extending to the computation of material properties and acoustic characteristics~\cite{schissler2017, tang2020}. Our Neural Acoustic Transfer leverages neural networks to represent acoustic transfer distributions.

\subsection{Monte-Carlo-based approach}
Recent advances in Monte Carlo (MC) methods have introduced novel solutions for boundary value problems, crucial in computer graphics and other domains. The Walk-on-Spheres (WoS) method, first introduced by Muller~\cite{muller1956} and later adapted to graphics by Sawhney and Crane~\cite{sawhney2020}, provides a robust and flexible approach for solving Dirichlet problems, leveraging MC ray tracing techniques~\cite{pharr2018}. These MC-based solutions demonstrate a favorable runtime-to-bias tradeoff and highlight the potential of Monte-Carlo techniques in numerical boundary value problem solutions.

\section{Neural Acoustic Transfer}
\label{sec:neural AT}

In this section, we will introduce our Neural Acoustic Transfer method. Moving beyond precomputed acoustic transfer methods, our neural-based approach accommodates conditional parameters, significantly enhancing its adaptability and relevance. This includes accommodating variable frequencies, object sizes, and object placements or rotation, thereby enabling nuanced and context-aware acoustic simulations. Once trained, our approach excels in prediction speed, allowing for fast exploration and experimentation within the acoustic condition space.

\subsection{Neural Representation of Acoustic Transfer Maps}

In this paper, we compute the acoustic transfer function within a small region containing objects that may vary in position, angle, or shape, and where sound vibration on the object surface can also vary. We explore methods to extend these solutions to external areas for sound rendering. The Complex-valued Far-Field Acoustic Transfer (FFAT) map, introduced by Chadwick et al.~\cite{shell-ffat}, represents the sound field around a stationary object. Additionally, Wang et al.~\cite{kleinpat} approximated the complex-valued FFAT map with a lightweight, real-valued expansion. Specifically, for a point $\mathbf{x}$ in space with spherical coordinates $(r, \theta, \phi)$, the FFAT map expresses the sound pressure at $\mathbf{x}$ as:
\begin{equation}
    |p(\mathbf{x})| = \sum_{i=1}^M \frac{\Phi_i(\theta, \phi)}{r^i} \ ,
\end{equation}
where the polynomials in $1/r$ describe direction-independent coefficients, and the function $\Phi_i$ captures the directional information of the radiating fields surrounding the object. $\Phi_i$ can be stored as a static image.

Such FFAT maps are computed and stored individually for each mode of an object, and the mode frequencies are fixed. However, in dynamic scenarios, such as when the frequency of a mode of a sound object changes (e.g., dynamically adjusting the object's material), when the frequency components of audio played by a sound source (speaker) change, or when the position and rotation of objects change, we need to adapt these maps. 

We propose neural FFAT Maps, where the values depend not only on the spherical coordinates \(\theta\), \(\phi\), and \(r\), but also on the surface vibration frequency \(f\) and other condition variables \(\mathbf{v} = (v_1, v_2, \ldots, v_n)\), which represent changes in position, angle, size, shape, etc. The neural FFAT Map can be written as:

\begin{equation}
    |p(\mathbf{x}, \mathbf{v}, f)| = \Phi(\theta, \phi, r, \mathbf{v}, f).
\end{equation}

\subsection{Problem Formulation}

In our approach, we utilize a neural network to represent our neural acoustic transfer maps. The neural network is defined by its parameters, including weights and biases, collectively denoted as \( \Theta \).

The inputs to our network are the listener's position \( \mathbf{x} \), conditional variable \(\mathbf{v} = (v_1, v_2, ..., v_n) \), and frequency variable $f$. The training process is optimizing the parameters \( \Theta \) to minimize the Mean Squared Error (MSE) loss across $N$ input-output pairs:
\begin{equation}
    \Theta^* = \underset{\Theta}{\mathrm{argmin}} \, \frac{1}{N} \sum_{i=1}^{N} \left( \Phi_{\Theta}(\theta_i, \phi_i, \mathbf{v}_i, f_i) - |p(\mathbf{x}_i,\mathbf{v}_i, f_i)| \right)^2,
\end{equation}
where \( \Theta^* \) represents the optimal set of parameters and $\Phi_{\Theta}(\theta_i, \phi_i, \mathbf{v}_i, f_i)$ is the output of neural network with input of $i$th data point.

As shown in \cref{fig:overview}, the network's prediction is given by:
\begin{equation}
    \Phi(\theta, \phi, \mathbf{v}, f) = \text{MLP} \left[\begin{array}{c}
         G(\theta, \phi, r) \\
         P(\mathbf{v}), \\
         P(f)
     \end{array}\right],
\end{equation}
where MLP denotes a multi-layer perceptron (composed of multiple fully connected layers and activation functions between them), and $P(\mathbf{v}), P(f)$ represents the positional encoding technique from NeRF~\cite{nerf}, which utilizes sinusoidal functions at varying frequencies for each dimension. To align with the positional encoding technique, we normalize each condition parameter within a consistent range of 0 to 1. For instance, for object movement, we constrain the object's movement within a bounding box, and the coordinates of the movement are normalized based on their relative position within the bounding box to a range between (0, 0, 0) and (1, 1, 1).

\subsection{Multi-resolution Feature Grid}
Our acoustic transfer modeling approach utilizes the multi-resolution
Hash grid, inspired by instant-NGP~\cite{mueller2022instant}, using \( G(\theta, \phi, r) \) to represent the multi-resolution Hash grid. Utilizing a purely MLP-based approach to fit sound field distributions can lead to a loss of detail, analogous to the limitations observed when fitting images with pure MLPs~\cite{mueller2022instant}. Consequently, the multi-resolution feature grid is employed to fit the complex neural FFAT maps. Utilizing spherical coordinates instead of Cartesian coordinates as the grid index enhances the capture of directional information of the radiating fields surrounding the objects. The multi-resolution feature grid involves the setup of multiple 3D lattice grids at various resolutions, encompassing the scene. Feature vectors at each point of these lattices are trilinearly interpolated to form a comprehensive feature representation at any given point, as described by the equation:
\begin{equation}
     G(\mathbf{x}) = \left[\begin{array}{c}
         \text{trilinear}\left(\mathbf{x}, V_0[\mathbf{x}]\right) \\
         \text{trilinear}\left(\mathbf{x}, V_1[\mathbf{x}]\right) \\
         \vdots \\
         \text{trilinear}\left(\mathbf{x}, V_{n-1}[\mathbf{x}]\right)
     \end{array}\right],
\end{equation}
where \( G: \mathbb{R}^3 \rightarrow \mathbb{R}^K \) represents the multi-resolution grid embedding function, and \( V_i[\mathbf{x}] \) denotes the \( K \)-dimensional feature vectors at the voxel corners enclosing \(\mathbf{x}\) on the \(i\)-th lattice. Hash encoding is incorporated to reduce memory costs~\cite{mueller2022instant}.

The values of $\theta$ and $\phi$ are first normalized from [-$\pi$, $\pi$] and [0, $\pi$] to [0, 1]. The frequency $f$ and condition variable $\mathbf{v}$ are also normalized according to their range before positional encoding.

\subsection{Network Design and Implementation}

In the neural network implementation, the MLP in neural FFAT maps consists of 4 hidden layers with 128 neurons per layer. ReLU activation is used after each layer except for the last. The input dimensions are adjusted to fit the outputs of positional encoding and grid encoding. The output dimension is 1 for scenarios where only a speaker produces sound, representing the sound pressure value at the listener point (only amplitude is considered in our experiments). For modal sound objects, the output dimension corresponds to the number of modes, representing the pressure value of each mode's vibration at the listener point.

The positional encoding comprises 6 frequency components, ranging from \( 2^0 \) to \( 2^5 \). Four levels of grid encoding are used, with resolutions ranging from \( 8 \times 8 \) to \( 64 \times 64 \), each level having a feature length of 4.

During the training phase, we optimize the neural network using the Adam optimizer~\cite{kingma2014adam}, starting with an initial learning rate of \( 1 \times 10^{-3} \). To enhance convergence and training stability, we implement a learning rate decay strategy, reducing the rate by a factor of 0.33 every one-third of the total training period. We implement the network and encoding modules with PyTorch~\cite{pytorch} and tiny-cuda-nn~\cite{tiny-cuda-nn}. All computations are conducted on a single Nvidia RTX 3080Ti GPU.

For data generation to train the neural network, we randomly sample \(\mathbf{v} = (v_1, v_2, \ldots, v_n)\). Since \(\mathbf{v}\) is normalized to the range 0-1, we use \(n\) random numbers from this range to sample \(\mathbf{v}\). With each sampled \(\mathbf{v}\), we construct the scene and obtain its surface triangle mesh. We then sample a random vibration frequency \(f\). In our experiments, the Neumann condition remains unchanged for each triangle. Given the frequency \(f\), the triangle mesh, and the Neumann condition, we use the Boundary Element Method (BEM) to solve for the unknown Dirichlet condition of this scene.

Next, we randomly sample \(\theta\), \(\phi\), and \(r\) within an enclosing sphere, with the radius constrained to be 1.5 to 3 times the size of the bounding box radius of the entire scene. The main bottleneck in data generation is solving the Dirichlet condition in BEM. However, computing the sound pressure at any point in space from the Neumann and Dirichlet conditions is much faster. Therefore, for each randomly sampled \(\mathbf{v}\) and \(f\), we generate 10,000 spatial sample points, obtaining their \(\theta\), \(\phi\), \(r\), and corresponding sound pressure as sample points.

Next, we will explain how we propose a faster Monte-Carlo-based BEM solution to further accelerate dataset generation for training.

\section{Monte-Carlo based Radiation Synthesis}
\label{sec:bem_solver}
To effectively compute the distribution of sound radiation under various scenario states \(\mathbf{v}\), an efficient numerical solution is indispensable. These scenarios often present challenges due to the variability of \(\mathbf{v}\), particularly the occurrence of singularities in conventional boundary integral equations at higher acoustic frequencies. To overcome these challenges, the Burton-Miller method is utilized for solving exterior Neumann boundary value problems. This approach incorporates a Hypersingular Boundary Integral Equation (HBIE) as follows:
\begin{equation}
\begin{aligned}
& \frac{1}{2} \phi(\boldsymbol{x}) - \int_{\Gamma} \phi(\boldsymbol{y}) \frac{\partial G(\boldsymbol{x}, \boldsymbol{y})}{\partial \boldsymbol{n}(\boldsymbol{y})} d S(\boldsymbol{y}) - \beta \int_{\Gamma} \phi(\boldsymbol{y}) \frac{\partial^2 G(\boldsymbol{x}, \boldsymbol{y})}{\partial \boldsymbol{n}(\boldsymbol{x}) \partial \boldsymbol{n}(\boldsymbol{y})} d S(\boldsymbol{y}) = \\
& -\int_{\Gamma} \partial_{\boldsymbol{n}} \phi(\boldsymbol{y}) G(\boldsymbol{x}, \boldsymbol{y}) d S(\boldsymbol{y}) - \beta \int_{\Gamma} \partial_{\boldsymbol{n}} \phi(\boldsymbol{y}) \frac{\partial G(\boldsymbol{x}, \boldsymbol{y})}{\partial \boldsymbol{n}(\boldsymbol{x})} d S(\boldsymbol{y}) - \frac{\beta}{2} \partial_{\boldsymbol{n}} \phi(\boldsymbol{y})
\end{aligned}
\label{eq:BurtonMiller}
\end{equation}
Here, \(\Gamma\) represents the surface of the scene, \(\phi(\boldsymbol{x})\) indicates the Dirichlet condition, and \(\partial_{\boldsymbol{n}} \phi(\boldsymbol{y})\) signifies the Neumann condition. The Green's function is denoted by \(G(\boldsymbol{x}, \boldsymbol{y})\), where \(\boldsymbol{n}(\boldsymbol{x})\) is the normal at the surface point \(\boldsymbol{y}\), and \(\beta = i / k\). Given the computational intensity due to the substantial number of variables in \(\mathbf{v}\) in complex scenarios, the open-source software \texttt{bempp}~\cite{bempp}, which is equipped with GPU acceleration, is recommended. Furthermore, we have implemented additional measures to further enhance the efficiency of the radiation solution process, addressing the need for accelerated computation in detailed and variable-rich environments.

\subsection{CUDA-accelerated BEM}
\label{CUDA-BEM}
To accelerate BEM computation on modern GPUs, we have reengineered the integration process of \texttt{bempp} utilizing CUDA, incorporating several significant improvements as follows.
We optimized the assembly process by directly constructing the matrices \(\mathbf{A}\) and \(\mathbf{B}\) in the linear equation \(\mathbf{A}\mathbf{g} = \mathbf{B}\mathbf{p}\), where \(\mathbf{g}\) and \(\mathbf{p}\) represent the discretized Neumann and Dirichlet conditions, respectively. This method contrasts with the original \texttt{bempp} implementation, which assembles a matrix for each individual term of the equation. This can significantly reduce redundant computations.
To enhance computational efficiency, particularly for integrations involving adjacent or identical elements, we employ a single CUDA block for each element-to-element integration. Within each block, threads are utilized for point-to-point internal Gaussian quadrature. This architecture markedly improves parallelism and minimizes latency associated with global memory access by strategically leveraging shared memory.
These improvements can optimize the use of hardware resources, leading to faster and more efficient solutions, averaging more than 20 times faster than the native \texttt{bempp} running on the same GPU.

\subsubsection{Monte-Carlo-based Integration Approximation}
For scenes where smooth Neumann conditions are common, we develop a Monte-Carlo integration-based BEM (BEM-MC) with singularity handling to solve the conventional boundary integral equation (i.e., \(\beta = 0\) in \cref{eq:BurtonMiller}) for faster training data synthesis. In this paper, we regard a scene with smooth Neumann conditions as one where the surface of the speaker has a Neumann condition of one and the surface of other passive objects has a Neumann condition of zero. Therefore, this MC based approximation is applied in such scenarios. 
For scenes with modal sound objects exhibiting high-frequency vibration modes, which we regard as having non-smooth Neumann conditions, this MC-based approximation is not used.

By applying Monte-Carlo integration to the conventional boundary integral equation and sampling \(M\) independent points \(\mathbf{y}_i\) on the boundary, we obtain:
\begin{equation}
    \frac{1}{2} p(\mathbf{x}) = \frac{1}{N}  \sum_{j=1}^{N} \frac{1}{q(\mathbf{y}_j)} \left( G(\mathbf{x}, \mathbf{y}_j) \frac{\partial p(\mathbf{y}_j)}{\partial n_j}  -  \frac{\partial G(\mathbf{x}, \mathbf{y}_j)}{\partial n_j} p(\mathbf{y}_j) \right),
    \label{eq:BIE}
\end{equation}
where \( q(\mathbf{y}_j) \) is the sample probability density on the domain boundary and \( n_j \) is the normal at boundary point \(\mathbf{y}_j\). For simplicity, we use uniform sampling, where \( \frac{1}{q(\mathbf{y}_j)} = |\Gamma| \), the overall area of the domain boundary. By evaluating the left-hand side at the same \( M \) sampling points \(\mathbf{y}_i\) (i.e., \( \mathbf{x} = \mathbf{y}_i \)), we derive a system of \( M \) equations:
\begin{equation}
    \frac{1}{2} p(\mathbf{y}_i) = \frac{1}{N}  \sum_{j=1}^{N} \frac{1}{q(\mathbf{y}_j)} \left( G(\mathbf{y}_i, \mathbf{y}_j) \frac{\partial p(\mathbf{y}_j)}{\partial n_j}  -  \frac{\partial G(\mathbf{y}_i, \mathbf{y}_j)}{\partial n_j} p(\mathbf{y}_j) \right),
    \label{eq:bem_linear_system}
\end{equation}
for \( i = 1, \ldots, M \). Solving this system of equations yields the values of \( p \) at the \( M \) sampling points, providing a discretized approximation of the Dirichlet boundary condition.

To address the singularity issue that may arise when \(\mathbf{y}_i = \mathbf{y}_j\), we derive an approximate formulation based on Green's function,
please refer to \cref{sec:address_singular} for technical detail. 

The main advantages of this technique include its straightforward random sampling approach, which serves as a simpler alternative to traditional remeshing in the BEM. It eliminates the reliance on and stringent requirements for mesh tessellation quality, a benefit that is also emphasized by other meshless methods~\cite{WoB,WoSt}. Additionally, the use of neural networks can significantly reduce the variance among multiple sampling results, thereby enhancing the method's reliability and computational efficiency. 
%Please refer to \cref{sec:mc_eval} for a detailed evaluation and analysis of our BEM-MC approach to show its high efficiency with moderate accuracy.

\subsection{Singularity Handling}
In the context of our Monte-Carlo-based approximation for BEM, a significant computational challenge arises when dealing with the Green's function and its gradient in integral calculations (\cref{eq:bem_linear_system}). The Green's function \( G(\mathbf{y}_i, \mathbf{y}_j) = \frac{e^{ik||\mathbf{y}_i-\mathbf{y}_j||}}{4\pi ||\mathbf{y}_i-\mathbf{y}_j||} \) has a denominator that includes the distance between points \( \mathbf{y}_i \) and \( \mathbf{y}_j \). When these points converge, Green's function approaches a singularity, leading to large values that can destabilize the computation and reduce precision.

The method's success generally hinges on precise sampling. To improve accuracy and efficiency, we use parallel Poisson disk sampling~\cite{bowers2010}, ensuring more uniform sample distribution and enhancing our acoustic transfer simulations.
Although Poisson disk sampling guarantees a minimum distance between different sampling points, singularities still occur when points in the integral equation coincide, i.e., \( \mathbf{y}_i = \mathbf{y}_j \). To address this, we divide the integral into two regions: a small disk centered at \( \mathbf{y}_i \) with radius \( r_0 \), and the rest of the boundary. Given the smoothness of the domain boundary, local approximations near \( \mathbf{y}_i \) are feasible and mathematically sound.

For regions outside the small disk, no singularity occurs, and the Monte-Carlo sampling method remains effective. The only necessary adjustment is a reduction in the sampling surface corresponding to the area of the small disk. However, within the small disk centered at \( \mathbf{y}_i \) with a radius of \( \epsilon \), special considerations are required to accommodate the singularity present in both the Green's function and its gradient.

In the context of solving \cref{eq:BIE} within the small disk region \( \Gamma_\epsilon \), we adopt a specialized approach to address the singularities in the integrals.

Firstly, consider the integral of the Green's function over \( \Gamma_\epsilon \). In polar coordinates, it is expressed as:
\begin{align}
    \int_{\Gamma_\epsilon} G(\mathbf{x}, \mathbf{y}) \frac{\partial p(\mathbf{y})}{\partial n_{y}} \, dy &= \int_{\Gamma_\epsilon} \frac{e^{ikr}}{4\pi r}  \frac{\partial p(\mathbf{y})}{\partial n_{y}} \, dy \\
    &= \int_0^{2\pi} \int_0^\epsilon \frac{e^{ikr}}{4\pi r}  \frac{\partial p(\mathbf{y})}{\partial n_{y}} r \, dr \, d\theta \\
    &= \int_0^{2\pi} \int_0^\epsilon \frac{e^{ikr}}{4\pi}  \frac{\partial p(\mathbf{y})}{\partial n_{y}} \, dr \, d\theta \ ,
\end{align}
where \( r \) denotes the Euclidean distance between the evaluation point \( \mathbf{x} \) and points \( \mathbf{y} \) on the disk. Applying Monte-Carlo integration with a single sample (\( N = 1 \)) at disk center $y_i$, we approximate the integral over \( \Gamma_\epsilon \) as:
\begin{equation}
    \int_{\Gamma_\epsilon} G(\mathbf{x}, \mathbf{y}) \frac{\partial p(\mathbf{y})}{\partial n_{y}} \, dy \approx 2\pi \epsilon \frac{1}{4\pi} \frac{\partial p(\mathbf{y}_i)}{\partial n_{i}} = \frac{\epsilon }{2} \frac{\partial p(\mathbf{y}_i)}{\partial n_{i}}.
\end{equation}

Next, we examine the integral involving the gradient of the Green's function over \( \Gamma_\epsilon \). Considering the perpendicularity of vector \( r \) to the surface normal $n_y$, this integral can be approximated as negligible:
\begin{equation}
    \int_{\Gamma_\epsilon} \frac{\partial G(\mathbf{x}, \mathbf{y})}{\partial n_y} p(\mathbf{y}) \, dy = \int_{\Gamma_\epsilon} -\frac{e^{i k r}}{4 \pi r^2}(1-i k r) \frac{\partial r}{\partial n_y} p(\mathbf{y}) \, dy = 0.
\end{equation}

Now, we can effectively address the singularities encountered in the integral computations within the small disk area.

\section{Evaluation of Neural Acoustic Transfer}

\begin{table*}[t]
\centering
\caption{Performance evaluation on scenes involving material and size editing of a plate with \textbf{60 vibration modes} (corresponding to \cref{fig:scale_comparison}). We compare the performance of BEM, NeuralSound, and our NAT for FFAT map computation for each material and size case. The average SNR and SSIM of FFAT maps across all modes, along with the computation times of each method for all 60 modes, are presented. The mesh used in BEM consists of 5174 triangles. The results underscore the superior accuracy and speed of our NAT.}
\begin{tabular}{c|ccc|ccc|ccc}
\hline
Case & \multicolumn{3}{c|}{BEM} & \multicolumn{3}{c|}{NeuralSound} & \multicolumn{3}{c}{NAT} \\
 & SNR$\uparrow$  & SSIM$\uparrow$  & Time$\downarrow$ & SNR$\uparrow$  & SSIM$\uparrow$  & Time$\downarrow$ & SNR$\uparrow$  & SSIM$\uparrow$  & Time$\downarrow$ \\
\hline
Small Metal & Inf & 1.0 & 11s & -4.43 & 0.22 & 0.05s & \textbf{12.68} & \textbf{0.85} & \textbf{0.002s} \\

Small Wood & Inf & 1.0 & 9s & -7.6 & 0.17 & 0.05s & \textbf{12.11}& \textbf{0.85}& \textbf{0.002s}\\

Small Ceramic &Inf &1.0 & 8s & -8.94 & 0.14 &0.05s &\textbf{10.39} &\textbf{0.83} & \textbf{0.002s}\\

Mid Ceramic &Inf &1.0 & 8s & 1.70 & 0.29 & 0.05s &\textbf{10.81} &\textbf{0.83} &\textbf{0.002s} \\

Large Ceramic &Inf & 1.0& 8s& 2.64& 0.20 & 0.05s& \textbf{9.86}& \textbf{0.81}& \textbf{0.002s}\\
\hline
\end{tabular}
\label{tab:performance_comparison}
\end{table*}

\begin{figure*}[ht]
\centering
\includegraphics[width=0.75\linewidth]{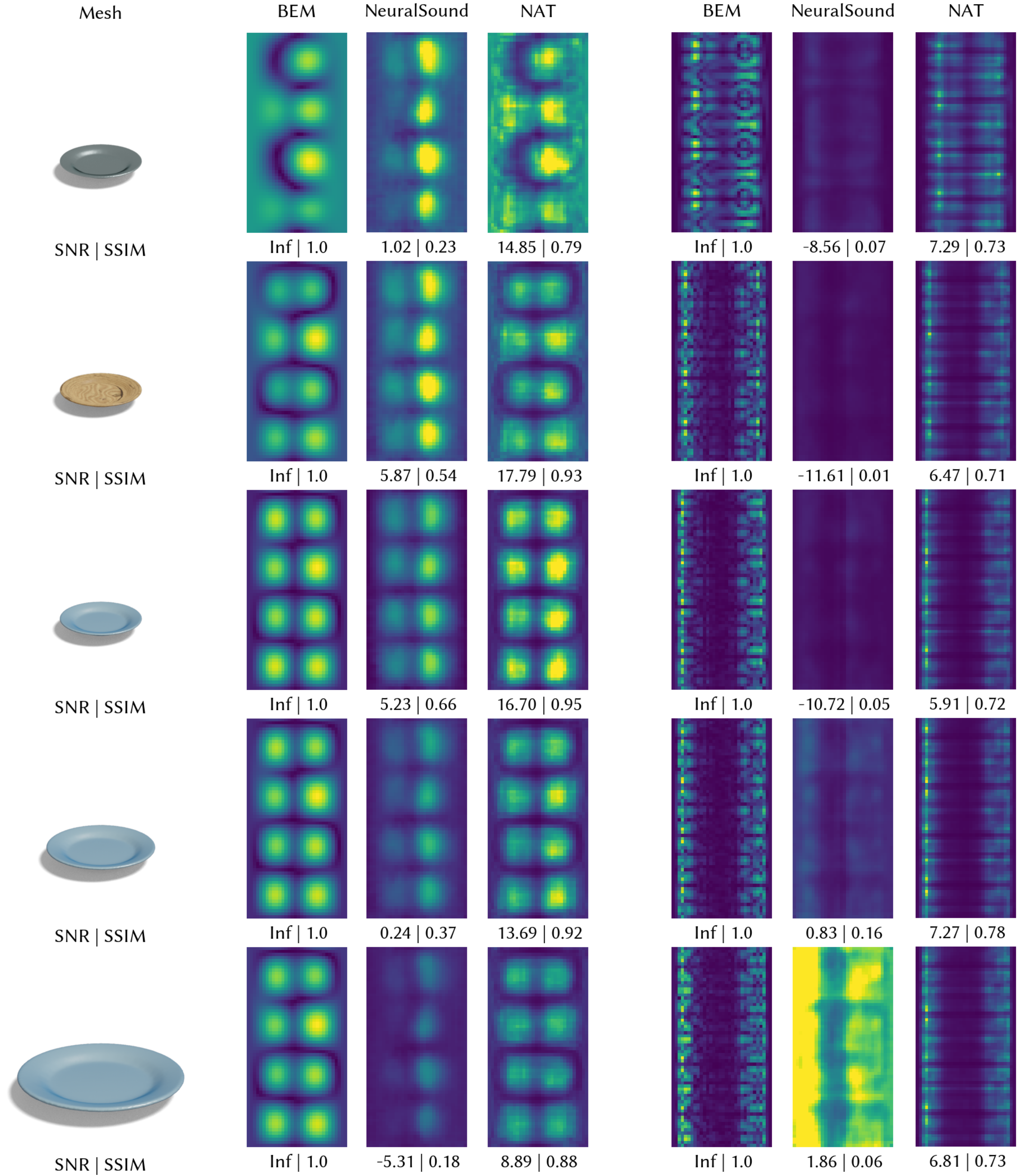}
\caption{Acoustic transfer tests on a scene involving material and size editing of a modal sound object with \textbf{60 vibration modes}. In each row, starting from the top, we display small metal, small wood, small ceramic, medium ceramic, and large ceramic plates, with the first and 50th modes shown for each. Our NAT demonstrates superior accuracy, closely matching the performance of BEM and significantly outperforming NeuralSound.}
\label{fig:scale_comparison}
\end{figure*}

\begin{figure*}[ht]
\centering
\includegraphics[trim={0cm, 0cm, 0cm, 0cm},clip,width=0.95\linewidth]{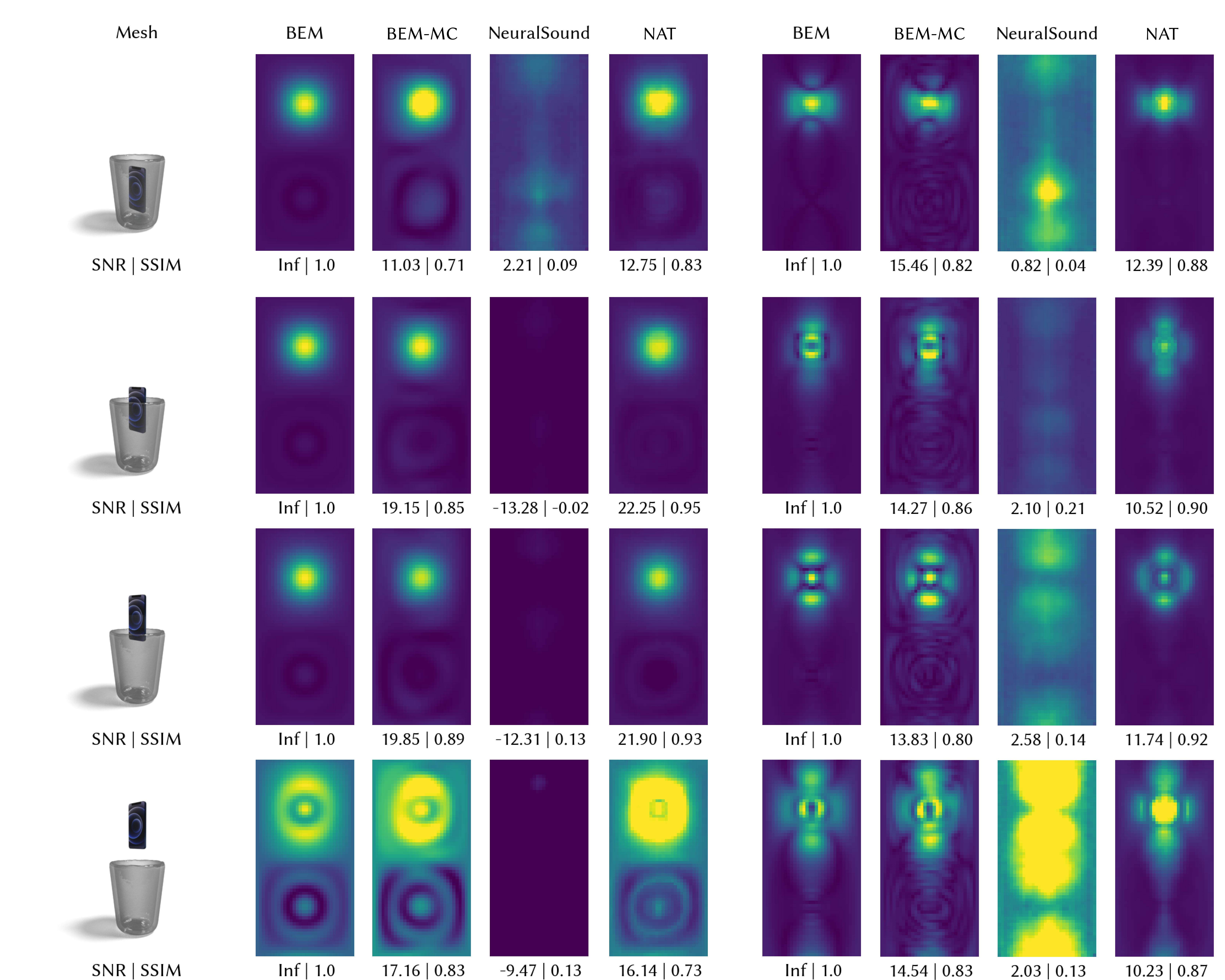}
\caption{Comparisons of FFAT maps for different relative positions of a cup and a vibrating phone at frequencies of 2000Hz (left four FFAT maps) and 7000Hz (right four FFAT maps). We highlight the superior accuracy of our NAT over NeuralSound. NeuralSound struggles with complex scenarios due to its simplified voxelization approach and focus on single objects. NAT often demonstrates better SSIM compared to BEM-MC, which generates its training dataset. This is because the neural network, when fitting data of similar scene conditions (i.e., similar relative positions of the cup and phone), effectively reduces the variance introduced by BEM-MC.}
\label{fig:audio_comparison}
\end{figure*}

% \subsection{Evaluation of  Neural Acoustic Transfer}
% \label{sec:experiment NAT}
We demonstrate the efficacy of Neural Acoustic Transfer through two application scenarios, showcasing its versatility and effectiveness in various acoustic modeling contexts. All the experiments are tested on a machine with single Nvidia RTX 3080Ti GPU.

\begin{figure*}[ht]
\centering
\includegraphics[width=0.9\linewidth]{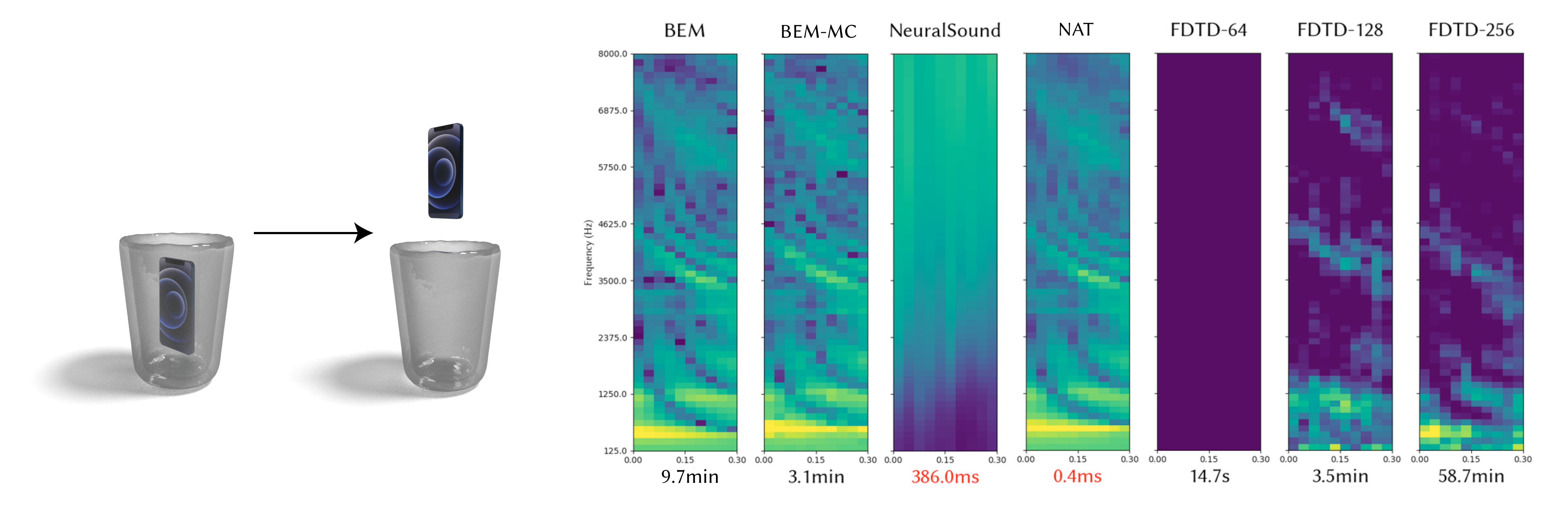}
\caption{A scenario where a vibrating phone, with only its bottom microphone position vibrating, is moved from inside a cup to a position 0.3m up of the cup. We measure the sound pressure at a point in front of the cup due to the phone's unit vibration at different frequencies. The figure on the right side illustrates this setup. Its horizontal axis represents the displacement of the moving phone, and the vertical axis denotes the frequency. We compared the results of seven methods in solving this problem. Among these, NAT and NeuralSound were significantly faster than the others. However, NeuralSound fails to accurately capture the variations of acoustic transfer.}
\label{fig:spectrogram}
\end{figure*}

\begin{table*}[ht]
\centering
\caption{Numerical comparison is conducted for the scenario where a phone playing sounds moves from the inside to the outside of a cup (corresponding to \cref{fig:audio_comparison}).
 For each frequency bin, every solver computes the FFAT maps for 80 cases. This computation incorporates 8 uniformly distributed frequencies within the frequency bin and 10 different phone positions. We present the computation time cost for \textbf{all 80 cases}, as well as the average SNR and SSIM of the FFAT maps for each method across these cases. NAT consistently remains the fastest among these methods.}
\begin{tabular}{c|ccc|ccc|ccc|ccc}
\hline
Frequncy Bin & \multicolumn{3}{c|}{BEM} & \multicolumn{3}{c|}{BEM-MC} & \multicolumn{3}{c|}{NeuralSound} & \multicolumn{3}{c}{NAT} \\
 & SNR$\uparrow$  & SSIM$\uparrow$  & Time$\downarrow$ & SNR$\uparrow$  & SSIM$\uparrow$  & Time$\downarrow$ & SNR$\uparrow$  & SSIM$\uparrow$  & Time$\downarrow$ & SNR$\uparrow$  & SSIM$\uparrow$  & Time$\downarrow$ \\
\hline
125Hz-2000Hz& Inf & 1.0 & 149s & 12.42 & 0.66 & 13.9s & -21.32 & 0.01 & 3.1s  & \textbf{12.38} & \textbf{0.69} & \textbf{0.06s} \\
2125Hz-4000Hz& Inf & 1.0 & 145s & 15.45 & 0.79 & 31.2s & -1.43 & 0.10 & 3.1s  & \textbf{17.35} & \textbf{0.90} & \textbf{0.06s} \\
4125Hz-6000Hz& Inf & 1.0 & 147s & 15.33 & 0.79 & 54.1s & 1.86 & 0.14 & 3.1s  & \textbf{15.24} & \textbf{0.90} & \textbf{0.06s} \\
6125Hz-8000Hz& Inf & 1.0 & 144s & 15.49 & 0.82 & 93.6s & 2.00 & 0.09 & 3.1s  & \textbf{11.26} & \textbf{0.86} & \textbf{0.06s} \\
\hline
\end{tabular}
\label{tab:audio_comparison}
\end{table*}

\subsection{Real-time Environmental Coupling Sound Effect}

In a dynamic scene, changes in the positions of objects lead to alterations in the gaps between them, thereby affecting the propagation of sound through these spaces. This results in a varying resonant cavity, producing distinct sound effects. To experiment with such effects, we use a scenario where a phone acts as a sound source (with only its bottom microphone position vibrating) and a cup that does not vibrate. The phone can freely move between the inside and outside of the cup. The movement of the phone is restricted to the y-axis, varying from 0 to 0.3m, with a frequency range for the phone's sound emission between 100-10000Hz.

We employ five different methods to solve for the acoustic transfer in this dynamic scene and perform both numerical and visual comparisons. These methods include BEM, BEM with Monte-Carlo-based approximation (BEM-MC) (see \cref{sec:bem_solver}), NeuralSound, Finite-Difference Time-Domain (FDTD) method, and our NAT. For a rich demonstration of the dynamic scene's sound effects generated using our method NAT, please refer to \cref{sec:extension} and the attached video.

The vertex count of all scenes is 10,004, and the triangle count is 20,000, which is used for the BEM solver. We employed BEM-MC to generate the dataset for NAT, sampling 10,000 points to compute the Monte-Carlo-based approximation.

Firstly, we employ BEM, BEM-MC, NeuralSound, and NAT to solve the FFAT maps for different cases. As highlighted in \cref{fig:audio_comparison}, NAT demonstrates superior accuracy over NeuralSound, showcasing its capability to accurately compute the spatial sound field distribution for various relative positions of objects. NeuralSound, however, struggles with complex scenarios due to its simplified voxelization approach and its focus on single objects. Even when we input the entire scene's information as a single object into NeuralSound, the results remain unsatisfactory.

NAT often exhibits better SSIM compared to BEM-MC, which produces the training data for NAT. This improvement is attributed to the neural network's ability to reduce the variance introduced by the MC based approximation in BEM-MC. For a detailed numerical comparison of the FFAT maps calculated by these methods, we present the results in \cref{tab:audio_comparison}. Our NAT method not only computes the fastest but also maintains high accuracy.

\begin{figure*}[ht]
\centering
\includegraphics[width=0.9\linewidth]{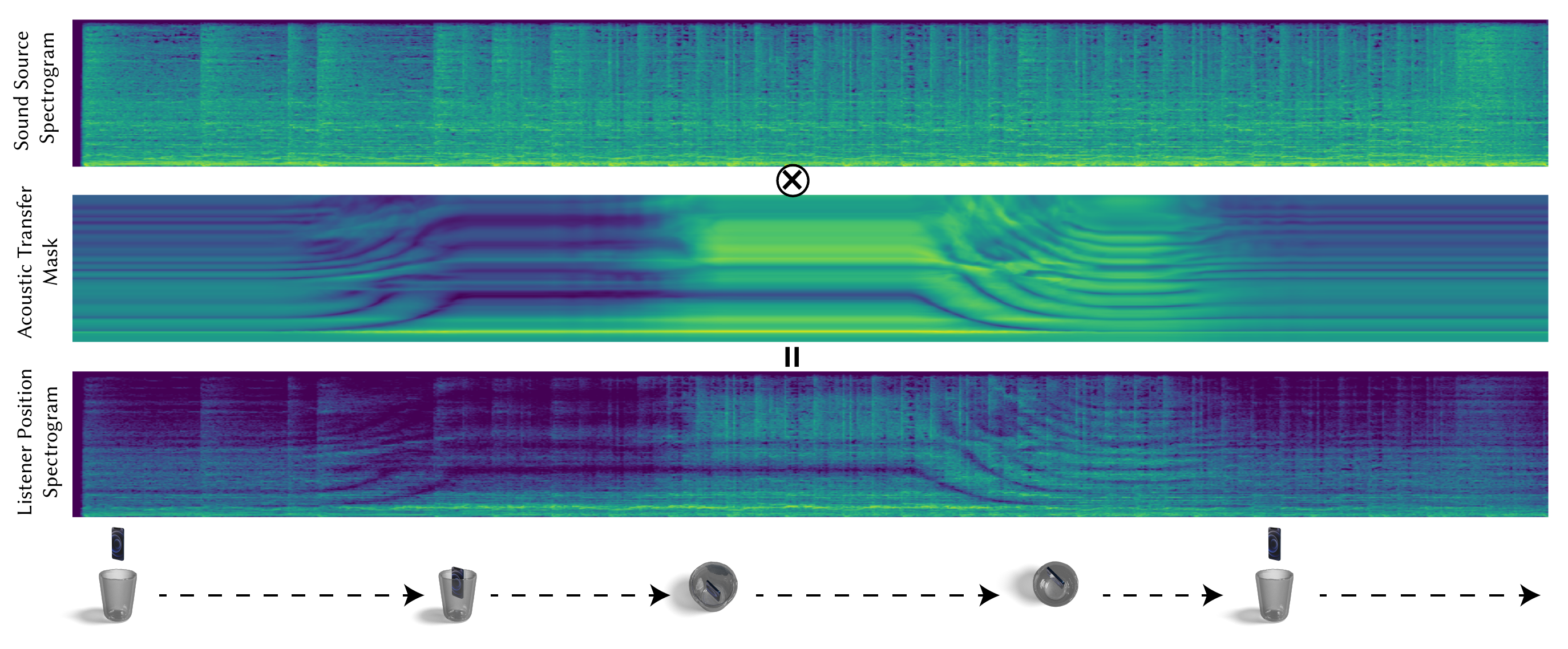}
\caption{Visualization demonstrating how our NAT synthesizes sound effects for a dynamic scene. Initially, we set an audio playback for the sound source (top spectrogram). Subsequently, NAT predicts the acoustic transfer for each frequency band at every moment, resulting in the acoustic transfer mask (middle spectrogram). Multiplying these two spectrogram yields the spectrogram corresponding to the audio at the listener's position (bottom spectrogram). It is evident that NAT captures variations in acoustic transfer caused by changes in the phone's position and listener (camera) position, leading to the synthesis of rich sound effects.}
\label{fig:extension_example}
\end{figure*}

\begin{figure*}[ht]
\centering
\includegraphics[width=0.95\linewidth]{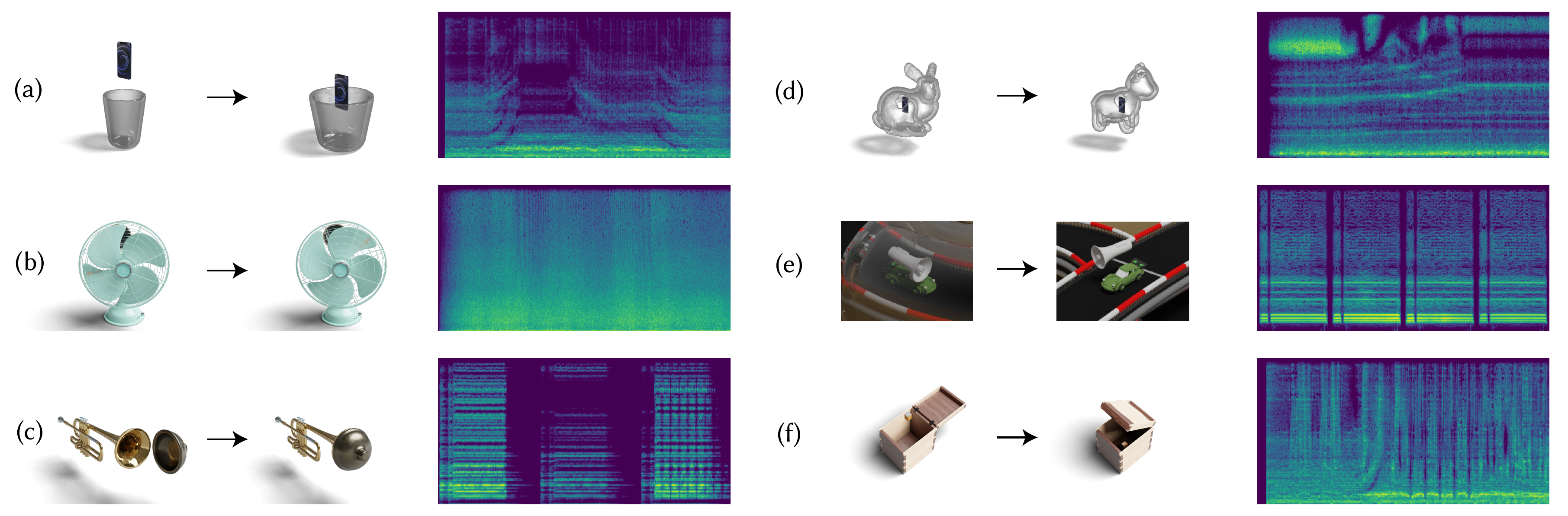}
\caption{Six extensive cases involve dynamic changes in acoustic transfer: (a) A phone playing audio moves in and out
of a cup. (b) A loudspeaker positioned behind a rotating fan. (c) A trumpet with a rotating lid covering its bell. (d) A phone positioned inside a shell while the geometric shape transforms from a bunny to a calf. (e) Two toy cars move on a racetrack that includes a transparent tunnel. (f) A rotating loudspeaker moves into a box with a cover that can also be rotated and resized.  The figure illustrates the dynamic changes of the object (left) and the listener position spectrogram (from top to bottom like \cref{fig:extension_example}) for each scene (right).}
\label{fig:extension}
\end{figure*}

In our next phase of testing, we apply BEM, BEM-MC, NeuralSound, NAT, and the Finite-Difference Time-Domain (FDTD) method~\cite{wang2018toward} to a scenario more reflective of practical applications. Rather than computing FFAT maps, our focus shifts to calculating the sound pressure values at a single listener position for various phone locations. The audio signal from a sound source can be decomposed into a finite number of frequency bins by transforming it into a spectrum. By calculating the acoustic transfer for each frequency bin, we are able to adjust the amplitudes of these frequency components in the source audio, leading to the synthesis of the final sound effect at the listener position. In this experiment, the frequency range of 0-8000 Hz is divided into 64 uniform frequency bins. We compute the sound pressure values at the listener position for each frequency bin, corresponding to different phone positions.

Given that FDTD is a time-domain solving algorithm, to ensure a fair comparison, we assigned a combined unit vibration signal consisting of 64 frequencies (centered on 64 frequency bins) to the phone's microphone. We then simulated the cup's movement from inside to outside over one second and calculated the signal at the same listener position. This was followed by a Fourier Transform to generate a frequency spectrum, thereby producing data comparable to the other four frequency-domain methods.

In the Finite-Difference Time-Domain (FDTD) framework, we opted for a first-order staircasing boundary handling scheme over Wang's second-order approach~\cite{wang2018toward} to avoid the substantial time and space costs required by the finer spatial resolution of the latter, as noted by Xue et al. \cite{xue2023improved}. For comparisons, we employed FDTD with different grid resolutions of \(64^3\), \(128^3\), and \(256^3\). The simulation time step was chosen to be as small as possible while adhering to the Courant-Friedrichs-Lewy (CFL) condition~\cite{bilbao2009finite}, aiming to minimize time costs. The simulation grid was designed to encompass a space three times the size of the bounding box of the scene objects. Additionally, Perfectly Matched Layers (PML)~\cite{liu1997perfectly} were implemented at the grid boundaries to minimize artificial reflections. We implemented FDTD using CUDA for a fair comparison.

As depicted in \cref{fig:spectrogram}, both NAT and NeuralSound exhibit superior computational speeds, with NAT surpassing NeuralSound by a considerable margin. However, NeuralSound fails to capture the propagation differences of various frequencies under different states, which is crucial for forming a resonant cavity. While FDTD is adept at managing dynamic scenarios, even with high-resolution grids (\(256^3\)), it falls short in capturing the intricate variations of each frequency at different phone positions, providing only a "low-resolution detail" of the changes. Additionally, as an offline algorithm, FDTD is far from achieving real-time speeds and necessitates recalculations upon alterations in object movements or sound source signals. Therefore, in computing the acoustic transfer for a dynamic scene, only our method NAT strikes a balance between speed and accuracy, offering the capability for real-time rendering.

\section{Validation of Monte-Carlo-based Approximation} 
\label{sec:mc_eval}
We begin with a comprehensive evaluation of the accuracy and efficiency of our Monte-Carlo-based approximation for BEM (BEM-MC). The analysis covers multiple aspects of performance, comparing our method with traditional techniques and investigating the impact of different sampling settings. 

\label{sec:address_singular}

\begin{figure*}[ht]
\centering
\includegraphics[width=0.9\linewidth]{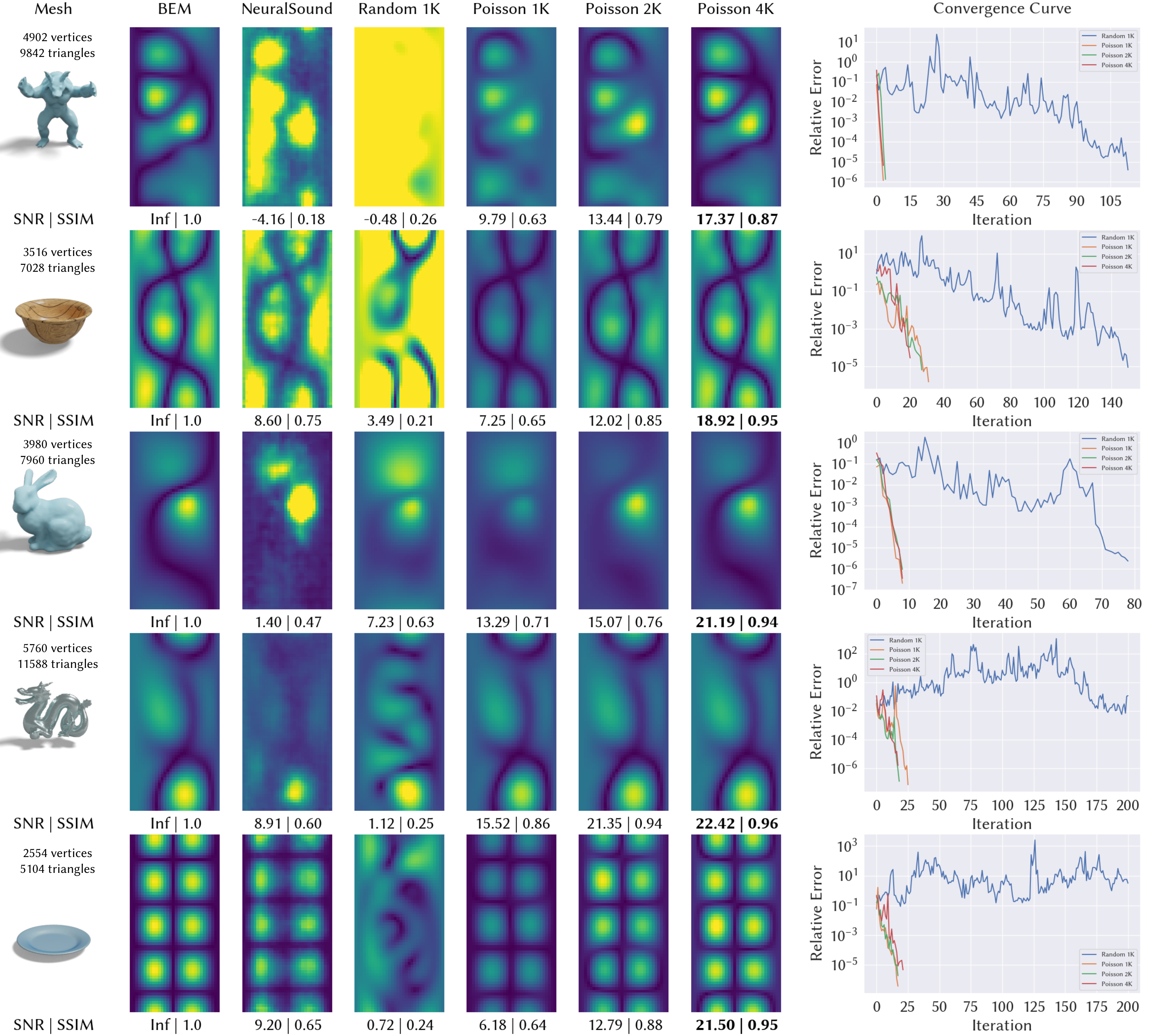}
\caption{Acoustic transfer was tested on diverse objects with different materials. For each object shown in the leftmost column, we use FFAT maps to visualize the results synthesized from different methods in the middle part, including BEM (ground truth), BEM-MC using random sampling (1K points), BEM-MC using Poisson disk sampling with different settings (1K, 2K, and 4K points), and NeuralSound \cite{NeuralSound}. We calculate SNR and SSIM for the mean value of the first 8 dominant modes. Poisson disk sampling demonstrates a clear advantage, particularly as the number of points increases. The rightmost column illustrates the convergence plot of our MC-based approximation using Poisson sampling, highlighting the effectiveness and stability of this strategy.}
\label{fig:ablation_study}
\end{figure*}

\subsection{Performance Analysis}

Our performance analysis for the BEM-MC employs the far-field acoustic transfer (FFAT) map as a key metric in evaluating the accuracy of acoustic transfer in modal vibrations with low frequency. BEM is employed as the ground truth. The performance of the tested solver is evaluated in terms of SNR and SSIM of FFAT maps.

This experiment involves the use of several 3D models and tests different sampling settings. We only compute the FFAT maps of the first 8 modes that exhibit smooth Neumann conditions. The performance of BEM-MC, both in terms of speed and accuracy, is directly influenced by the sampling strategy employed, particularly the number of sample points. Therefore, we examine various configurations of Poisson disk sampling, each with a different number of sample points. Given that Poisson disk sampling inherently increases the computational time for sampling, it is crucial to demonstrate that the accuracy gains provided by Poisson disk sampling can fully offset the additional computational cost. To this end, we also compare it with random sampling. For all sampling strategies, the settings for the linear solver are consistent: a tolerance of \(1 \times 10^{-6}\) and a maximum of 200 iterations.

\begin{table}[th]
\centering
\caption{The time cost (in seconds) for different methods solving the acoustic transfer maps for the first \textbf{8 modes} of various testing modal sound objects, corresponding to the testing objects shown in \cref{fig:ablation_study}, is presented. BEM-MC shows a computational time cost comparable to NeuralSound while achieving significantly improved accuracy. Compared to CUDA-accelerated BEM, our method demonstrates a substantial advantage in speedup, with all tests conducted on the same GPU.}
\scalebox{0.9}{
\begin{tabular}{l|ccccc}
\hline
\textbf{Runtime (s) } & \textbf{Armadillo} & \textbf{Bowl} & \textbf{Bunny} & \textbf{Dragon} & \textbf{Plate} \\
\hline
NeuralSound & 0.06 & \textbf{0.06} & 0.07 & \textbf{0.06}& \textbf{0.05} \\
Random 1K & 0.08 & 0.10 & 0.06 & 0.13 & 0.13 \\
Poisson 1K & \textbf{0.03} & \textbf{0.05} & \textbf{0.03} & \textbf{0.04} & \textbf{0.04} \\
Poisson 2K & \textbf{0.04} & 0.09 & \textbf{0.05} & 0.07& 0.07 \\
Poisson 4K & 0.08 & 0.28 & 0.11 & 0.16 &  0.15 \\
BEM & 2.2 & 1.2 & 1.2 & 2.7 &  1.1 \\
\hline
\end{tabular}
}
\label{tab:computation_times}
\end{table}

\begin{figure}[ht]
\centering
\includegraphics[width=\linewidth]{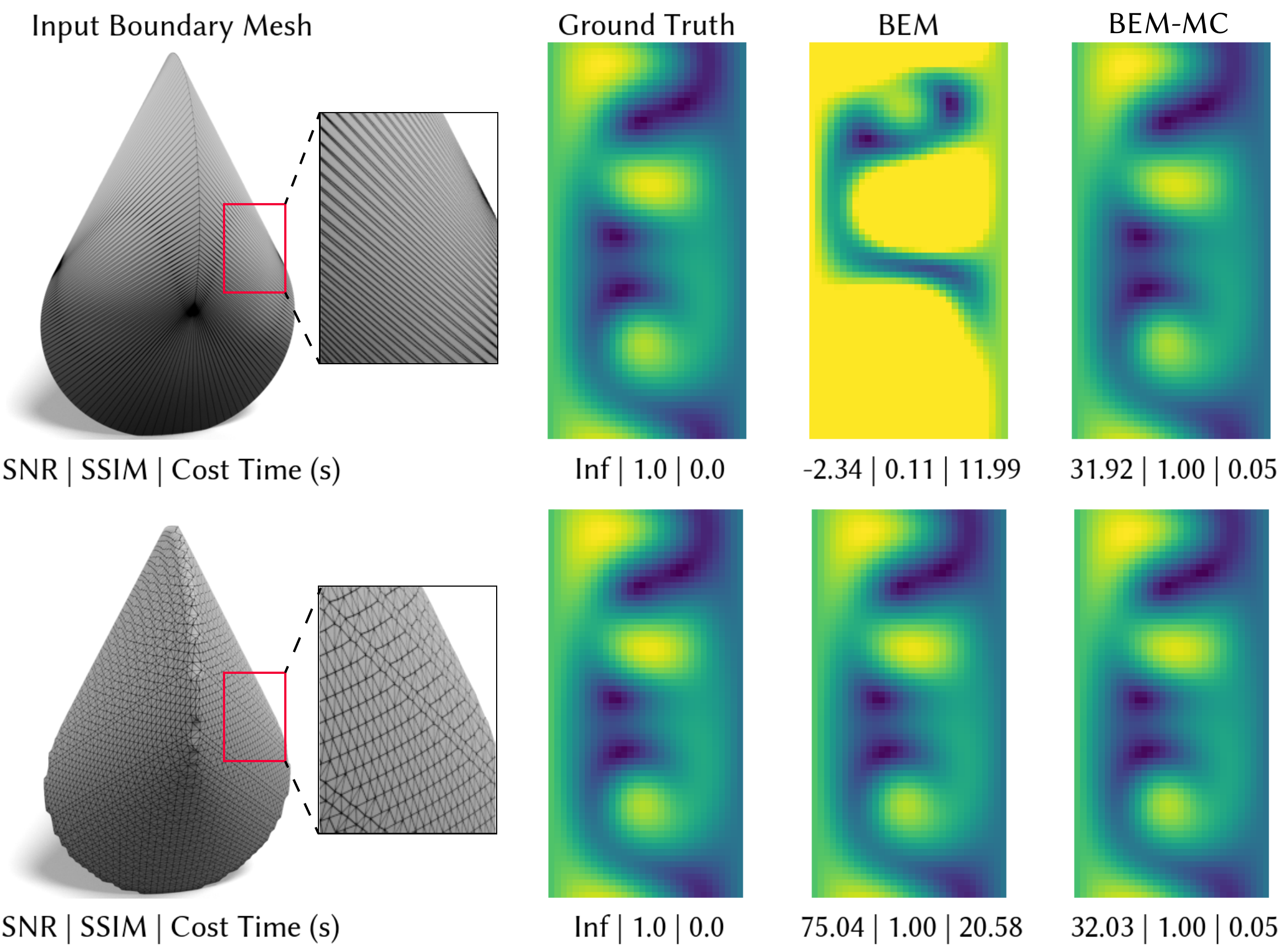}
\caption{Analytical test case involves three dipole sound sources positioned at different y-axis locations within a mesh to compare acoustic modeling using BEM-MC and BEM. We use both irregular mesh (upper row, with sliver triangles) and regular mesh (lower row) for evaluation. The FFAT map results illustrate obvious errors in BEM with irregular mesh elements, attributed to local aliasing of boundary data, while BEM-MC maintains high accuracy and is not sensitive to the mesh quality.}
\label{fig:bem_comparison}
\end{figure}

In \cref{fig:ablation_study}, we present the acoustic transfer results of various 3D models. BEM-MC with Poisson disk sampling exhibits very close results to the ground truth while BEM-MC with random sampling have very low accuracy in terms of SNR and SSIM. The study clearly demonstrates a substantial increase in precision when utilizing Poisson disk sampling as compared to random sampling. This improvement is attributed to the fact that Poisson disk sampling provides more uniformly distributed points, leading to less variance in equation solving. Furthermore, while Poisson disk sampling necessitates additional initial time investment (around 10 ms) compared to random sampling (around 1 ms), it offers a significantly faster convergence rate. Detailed time costs for these methods are provided in \cref{tab:computation_times}, further illustrating the efficiency and effectiveness of our approach in acoustic modeling.

\subsection{Accuracy Evaluation}
BEM-MC has advantages over traditional methods in terms of mesh robustness, accuracy, and computational efficiency.

\textbf{Comparison with BEM:} The resilience of BEM-MC to mesh quality represents an advantage, particularly in scenarios involving suboptimal mesh structures, especially in data from real-world collection. In \cref{fig:bem_comparison}, we compare BEM-MC with BEM (CUDA-accelerated version as described in Sec. \ref{CUDA-BEM}) using analytical test cases as suggested in~\cite{sigcourse}. BEM-MC consistently outperforms in challenging mesh conditions, where traditional BEM often exhibits significant errors. This advantage arises from BEM's susceptibility to local aliasing of boundary data, particularly in meshes with irregular elements. In contrast, BEM-MC is less dependent on mesh quality, as point sampling is not sensitive to mesh conditions. As shown in \cref{fig:ablation_study} and \cref{tab:computation_times}, BEM-MC surpasses the CUDA-accelerated BEM. 

\begin{table*}[t]
\centering
\caption{Performance evaluation on scenes involving materials and size editing of a modal sound object with the first \textbf{8 modes} (corresponding to \cref{fig:scale_comparison_8}). We compare the performance of BEM, NeuralSound, BEM-MC, and NAT (trained with data generated from BEM-MC) for FFAT map computation for each material and size case. The average SNR and SSIM of FFAT maps across the first 8 dominant modal frequencies, along with the computation times for each method, are presented. The results underscore the superior accuracy and speed of our NAT across various material and size contexts.}
\begin{tabular}{c|ccc|ccc|ccc|ccc}
\hline
Case & \multicolumn{3}{c|}{BEM} & \multicolumn{3}{c|}{BEM-MC} & \multicolumn{3}{c|}{NeuralSound} & \multicolumn{3}{c}{NAT} \\
 & SNR$\uparrow$  & SSIM$\uparrow$  & Time$\downarrow$ & SNR$\uparrow$  & SSIM$\uparrow$  & Time$\downarrow$ & SNR$\uparrow$  & SSIM$\uparrow$  & Time$\downarrow$ & SNR$\uparrow$  & SSIM$\uparrow$  & Time$\downarrow$ \\
\hline
Small Metal & Inf & 1.0 & 1.5s  & 8.37 & 0.83 & 0.15s & 6.77 & 0.52 & 0.05s & \textbf{11.79} & \textbf{0.93} & \textbf{0.002s} \\

Small Wood & Inf & 1.0 & 1.2s  & 6.63& 0.85& 0.15s & 5.52& 0.49 & 0.05s & \textbf{10.73}& \textbf{0.93}& \textbf{0.002s}\\

Small Ceramic &Inf &1.0 &1.1s  &6.25 &0.84 &0.15s& 6.17 &0.54 &0.05s &\textbf{9.57} &\textbf{0.92} & \textbf{0.002s}\\

Mid Ceramic &Inf &1.0 &1.1s  &4.14 &0.78 &0.14s &2.92 &0.30 &0.05s &\textbf{7.28} &\textbf{0.88} &\textbf{0.003s} \\

Large Ceramic &Inf & 1.0& 1.1s& 2.24& 0.72& 0.14s& -0.66& 0.17& 0.05s &  \textbf{5.11}& \textbf{0.82}& \textbf{0.002s}\\
\hline
\end{tabular}
\label{tab:performance_comparison_8}
\end{table*}
\begin{figure*}[ht]
\centering
\includegraphics[width=0.8\linewidth]{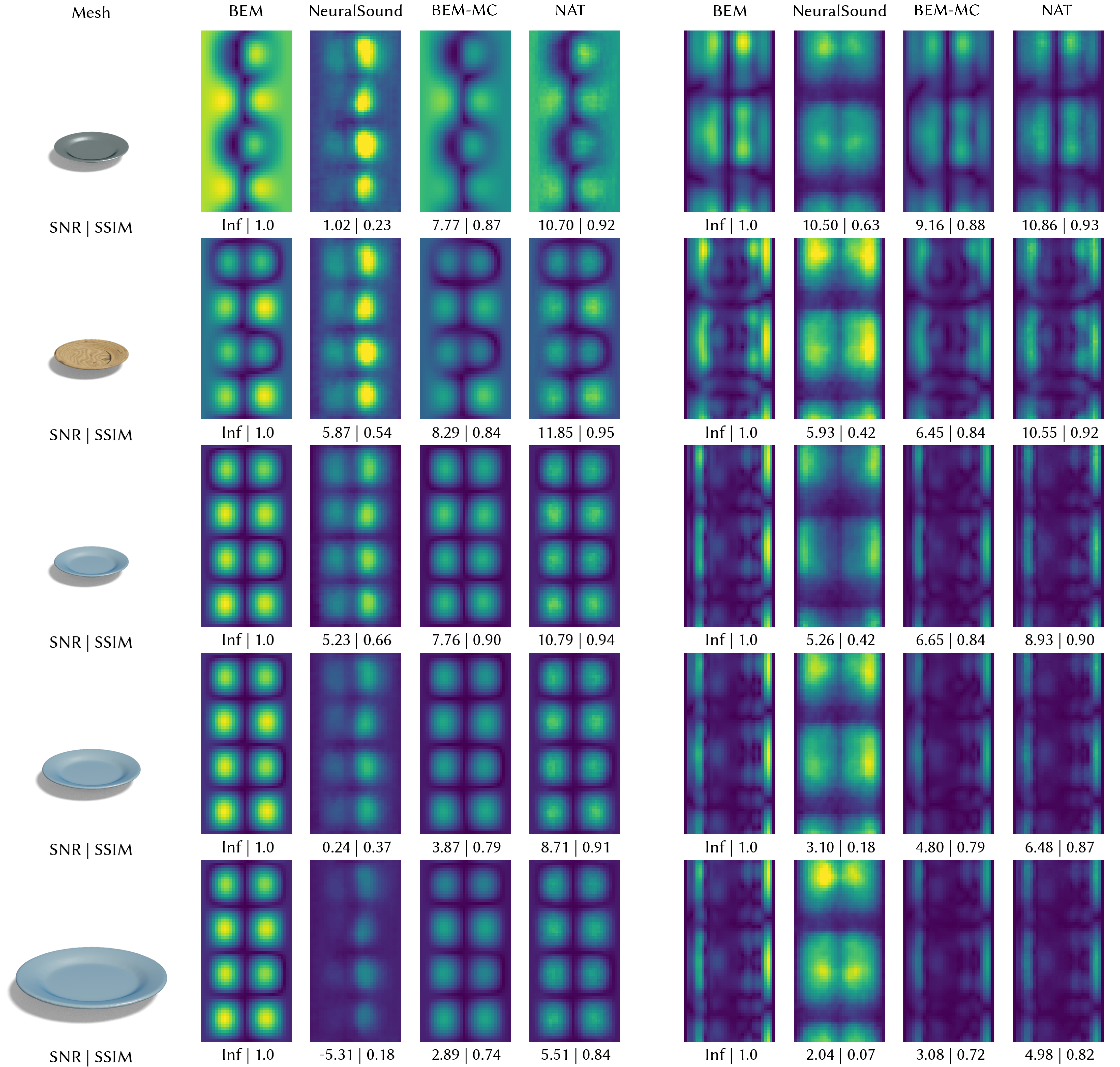}
\caption{Acoustic transfer tests on a scene involving materials and size editing of a modal sound object with the first \textbf{8 vibration modes}. In each row, starting from the top, we display small metal, small wood, small ceramic, mid ceramic, and large ceramic plates, with two random modes shown for each. Both BEM-MC and NAT demonstrate superior accuracy, closely matching the performance of BEM and significantly outperforming NeuralSound.}
\label{fig:scale_comparison_8}
\end{figure*}

\subsection{Dataset Synthesis and Validation} 
\label{sec:material_size_edit}

For this evaluation, we selected a plate as our testing object and varied its material properties by adjusting the ratio of Young's modulus and density within the range of \(7.8 \times 10^6\) to \(2.6 \times 10^7\), encompassing common materials like glass, ceramic, wood, and metal. Additionally, the diameter of the plate was varied randomly between 0.1 m and 0.2 m. The tested solver was required to compute the FFAT maps of the first 60 modes after changes in material or size. As can be observed from the Helmholtz equation, the value of frequency multiplied by size determines the resulting acoustic transfer, so we set the frequency multiplied by size as the condition parameter \(v\).

For training NAT, we utilized the CUDA-accelerated BEM to serve as the ground truth, and solved the acoustic data of the first 60 dominant modes of 1,000 plates, each with a randomly assigned material and size. Due to the complexity and non-smoothness of the high-frequency mode shapes, we did not use MC based approximation for maintaining numerical stability. However, we found that for low-frequency mode shapes, our MC based approximation remains stable. For experimental results and analysis of NAT trained with data from the first 8 dominant modes precomputed by BEM using our MC based approximation, please refer to \cref{sec:scale_8}.

We compare three methods: BEM (ground truth), NeuralSound~\cite{NeuralSound}, and our NAT, using Signal-to-Noise Ratio (SNR) and Structural Similarity Index (SSIM) of FFAT maps as metrics. NeuralSound~\cite{NeuralSound} is a neural network architecture designed for fast modal sound synthesis. In this paper, we only consider the acoustic transfer part of NeuralSound, which encodes surface displacement and frequency of vibration modes into scalar-valued FFAT maps. These maps compress the acoustic transfer function for sound rendering. The comparative FFAT maps of all the tested solvers in this evaluation are depicted in \cref{fig:scale_comparison}. Our NAT consistently exhibits closer accuracy to BEM compared to NeuralSound.

The comparison of time cost and average precision for solving the FFAT maps of the first 60 modes across different solvers is presented in \cref{tab:performance_comparison}. This table clearly demonstrates the significant advantages of our NAT in terms of both speed and accuracy over other methods. NAT is sufficiently fast, eliminating the need for pre-computing FFAT maps before real-time evaluation, unlike other methods. Given that our FFAT map resolution is \(64 \times 32\), the computational cost for NAT is only 2ms. This implies that NAT can predict acoustic transfer at 500 FPS for 1024 positions simultaneously. NAT can serve as a high-performance, accurate, and adaptable precomputed acoustic transfer method for modal sound objects, allowing for real-time material and size editing. Please refer to the attached video for the demonstration.

\subsubsection{Real-time Material and Size Editing for Modal Sound Object with BEM-MC data generation}
\label{sec:scale_8}

In this section, we conduct experiments similar to those in \cref{sec:material_size_edit}, but here we focus only on fitting the first 8 dominant modes of the plate and use BEM-MC to generate data for NAT. We compare four methods: BEM (ground truth), NeuralSound~\cite{NeuralSound}, BEM-MC, and NAT (trained with data generated from BEM-MC). The comparative FFAT maps of the four tested solvers in this evaluation are depicted in \cref{fig:scale_comparison_8}. Both BEM-MC and NAT consistently exhibit closer accuracy to BEM compared to NeuralSound. It is observed that the accuracy of NAT often rivals or even surpasses that of BEM-MC, which is employed for training NAT. This enhanced accuracy of NAT can be attributed to the neural network's ability to reduce the variance caused by the limited number of sampling points in BEM-MC. The comparison of time cost and average precision (SNR and SSIM) for solving the FFAT maps of the first eight modes across different solvers is presented in \cref{tab:performance_comparison}.

\section{Extensive Cases}
\label{sec:extension}

In our extended experiment, we employ NAT to address the sound effects in six complex scenarios, each involving dynamic changes in the sound field.

We first briefly outline how to synthesize the corresponding acoustic effects for a dynamic scene using our NAT. Initially, we need to train NAT based on data from BEM-MC for this dynamic scenario. Subsequently, as illustrated in \cref{fig:extension_example}, we can set an audio playback for the sound source, obtain its spectrogram (sound source spectrogram), and then manipulate the placement of objects and camera positions within the scene arbitrarily. Utilizing NAT, we acquire the acoustic transfer for each time frame under each frequency band at the listener position. This results in an acoustic transfer mask with the same size as the spectrogram of the sound source. Multiplying this mask with the original spectrogram yields the spectrogram of the sound at the listener position, allowing us to synthesize the corresponding audio. As depicted in \cref{fig:extension_example}, the masks generated by NAT exhibit rich details, corresponding to the variations in the positions of the phone and camera in the animation. For a more comprehensive understanding, please refer to the accompanying video for the sound effects synchronized with the animation.

In addition to the aforementioned dynamic scenario, the spectrograms of the sound source, masks, and listener position spectrograms for the other three similar dynamic scenes are presented in \cref{fig:extension}. We create four scenes to examine our NAT.

\begin{itemize}
    \item \textbf{Phone and Cup:} A phone playing audio moves in and out of a cup, and the cup size changes. The position of the phone from 0m to 0.2m and the width size of the cup from 0.05m to 0.1m are regarded as condition variables.
    \item \textbf{Rotating Fan:} A fan is rotating, and a loudspeaker as the sound source is behind the fan. The rotation from 0 to \(2\pi\) is regarded as the condition variable.
    \item \textbf{Trumpet with Rotating Lid:} A trumpet with a rotating lid covering its bell, with the sound source placed inside its bell. The rotation from 0 to \(\pi\) is regarded as the condition variable.
    \item \textbf{Morphing Shape:} A phone (sound source) positioned inside a shell while the geometric shape transforms from a bunny to a calf, affecting the acoustic transfer. The shape parameter from 0 (bunny) to 1 (calf) is regarded as the condition variable.
 \end{itemize}   
 
\subsection{Multiple Dynamically-Coupled Cases}
Furthermore, we test our NAT using multiple dynamically-coupled scenarios, which are highly complex situations in which many elements are involved dynamically in changing the radiation fields simultaneously. 
 \begin{itemize}   
    \item \textbf{Rotating Loudspeaker in Box:} A rotating loudspeaker moves into a box with a cover that can also be rotated and resized. The rotation from 0 to \(2\pi\) of the loudspeaker, its height from 0m to 0.2m, the rotation of the box cover from 0 to \(\pi\), and the width size of the box from 0.15m to 0.3m are all regarded as condition variables.
    The complexity of this scene lies in the rotation and movement of the sound source, which means the orientation of sound transfer is dynamically changing. Additionally, the relative positions of the sound sources within their environment (a box) and the continuous opening and closing of the box's lid indicate that the sound field coupling with the environment is also continuously changing.
    \item \textbf{Toy Cars on Racetrack:} Two toy cars move on a racetrack that includes a transparent tunnel. The cars play different sounds and pass through the tunnel. The position of the two cars from 0 (start point) to 1 (end point), the different volumes of the two cars from 0 to 1, and the tunnel's length (represented by 0 to 1 for shortest to longest states) are regarded as condition variables. 
    The complexity of this scenario is highlighted by the presence of two dynamically moving sound sources, and a tunnel whose length changes dynamically and is coupled with the sound field.
\end{itemize}

Through rigorous testing in these dynamic scenarios, our NAT demonstrates the capability to accurately capture variations in the sound field resulting from different scene changes. The scene information and corresponding model data for NAT across these all scenarios are summarized in \cref{tab:extension}. Examining the data in the table reveals that NAT necessitates only a few hours of precomputation and training on a 
GPU to achieve fully real-time and interactive synthesis of sound in dynamic scenes. Notably, the computation of the mask   for half a minute of audio takes merely 3 milliseconds. Furthermore, the size of the network model, once precomputed and trained, is remarkably small. This efficiency in both computational requirements and model storage highlights the effectiveness of our approach.

\begin{table*}[h]
\centering
\caption{Audio Scene Characteristics and Model Details: Our NAT demonstrates the ability to complete precomputation (using BEM with MC-based approximation for Neumann boundary problems corresponding to 10,000 - 20,000 scene condition setups) and training within a few hours, requiring minimal storage space (around 1 megabyte). Subsequently, it can provide an acoustic transfer mask for the sound of a dynamic scene lasting half a minute with an extremely small time cost, on the order of a few milliseconds.}
\begin{tabular}{lccccccc}
\hline
\textbf{Scene}     & \textbf{\begin{tabular}[c]{@{}c@{}}Audio \\ Length\end{tabular}} & \textbf{\begin{tabular}[c]{@{}c@{}}Num of\\ Condition\end{tabular}} & \textbf{\begin{tabular}[c]{@{}c@{}}AT Mask \\ Resolution\end{tabular}} & \textbf{\begin{tabular}[c]{@{}c@{}}Precompute \\ Time\end{tabular}} & \textbf{\begin{tabular}[c]{@{}c@{}}Training \\ Time\end{tabular}} & \textbf{\begin{tabular}[c]{@{}c@{}}Model \\ Size\end{tabular}} & \textbf{\begin{tabular}[c]{@{}c@{}}Inference \\ Time\end{tabular}} \\ \hline
Phone \& Cup       & 28 s                                                             & 2                                                                   & 512 * 840                                                              & $\sim$1 h                                                                 & $\sim$5 mins                                                            & 0.4 MB                                                         & 3 ms                                                               \\
Rotating Fan       & 26 s                                                             & 1                                                                   & 512 * 840                                                              & $\sim$2 h                                                                 & $\sim$5 mins                                                            & 0.8 MB                                                         & 3 ms                                                               \\
Trumpet            & 11 s                                                             & 1                                                                   & 512 * 330                                                              & $\sim$2 h                                                                 & $\sim$5 mins                                                            & 0.8 MB                                                         & 2 ms                                                               \\
Morphing Shape     & 7 s                                                              & 1                                                                   & 512 * 210                                                              & $\sim$1 h                                                                 & $\sim$5 mins                                                            & 0.8 MB                                                         & 1 ms                                                               \\
Toy Car            & 40 s                                                              & 5                                                                   & 512 * 1200                                                             & $\sim$1 h                                                                 & $\sim$5 mins                                                            & \multicolumn{1}{l}{0.6 MB}                                     & 4 ms                                                               \\
Loudspeaker in Box & 20 s                                                              & 4                                                                   & 512 * 600                                                              & $\sim$1 h                                                                 & $\sim$5 mins                                                            & 1.2 MB                                                         & 3 ms                                                                \\ \hline
\end{tabular}
\label{tab:extension}
\end{table*}

\section{Conclusion, Limitations, and Future Work}

We presented the Neural Acoustic Transfer framework, enabling real-time interactions in complex acoustic environments. Through extensive demonstrations, our approach can be an adaptable and powerful tool for diverse interactive acoustic modeling scenarios. The versatility and real-time processing capabilities of our method make it an advancement in the field of acoustic simulation. In addition, by leveraging Monte-Carlo-based approximation for synthetic training data, our approach significantly accelerates data production while maintaining sufficient accuracy for sound effect simulation. This represents a promising and valuable endeavor.

Despite its advancements, our method exhibits certain limitations.  The hyperparameters of the neural network within NAT have not undergone fine-tuning. Exploring the optimal selection of encoders and configurations for hyperparameters represents a valuable avenue for future research.
In addition, the current conditions for fitting the neural network to the scene parameters are relatively limited. Introducing a more extensive set of scene parameters and conducting effective and rapid training on these parameters is also a crucial aspect for future work. The inclusion of a larger variety of scene conditions in the training process holds significant promise for enhancing the versatility and performance of NAT in capturing the intricacies of diverse dynamic scenarios.

An exciting avenue for future exploration is the application of our framework in Virtual Reality (VR) environments. The real-time, dynamic capabilities of our approach hold significant potential for creating immersive and complex acoustic scenarios within VR settings. This integration could pave the way for new interactive experiences and advancements in the field of sound synthesis in virtual environments.

\bibliographystyle{IEEEtran}
\bibliography{bibliography}

\end{document}